# Reliable Feature Selection for Adversarially Robust Cyber-Attack Detection

João Vitorino[0000-0002-4968-3653], Miguel Silva[0009-0008-6630-9939], Eva Maia[0000-0002-8075-531X] and Isabel Praça[0000-0002-2519-9859]

Research Group on Intelligent Engineering and Computing for Advanced Innovation and Development (GECAD), School of Engineering, Polytechnic of Porto (ISEP/IPP), 4249-015 Porto, Portugal
`{jpmvo,mdgsa,egm,icp}@isep.ipp.pt`

**Abstract.** The growing cybersecurity threats make it essential to use high-quality data to train Machine Learning (ML) models for network traffic analysis, without noisy or missing data. By selecting the most relevant features for cyber-attack detection, it is possible to improve both the robustness and computational efficiency of the models used in a cybersecurity system. This work presents a feature selection and consensus process that combines multiple methods and applies them to several network datasets. Two different feature sets were selected and were used to train multiple ML models with regular and adversarial training. Finally, an adversarial evasion robustness benchmark was performed to analyze the reliability of the different feature sets and their impact on the susceptibility of the models to adversarial examples. By using an improved dataset with more data diversity, selecting the best time-related features and a more specific feature set, and performing adversarial training, the ML models were able to achieve a better adversarially robust generalization. The robustness of the models was significantly improved without their generalization to regular traffic flows being affected, without increases of false alarms, and without requiring too many computational resources, which enables a reliable detection of suspicious activity and perturbed traffic flows in enterprise computer networks.

## 1     Introduction

The growing quantity and complexity of cybersecurity threats is making it essential for organizations of all sizes to improve the protection of their digital assets [1]. Network intrusion detection systems monitor the traffic of an enterprise computer network and identify potentially harmful behavior that could threaten the integrity, confidentiality, or availability of computer resources [2]. These systems can use Artificial Intelligence (AI), and more specifically Machine Learning (ML) models, to automatically perform network traffic analysis and anomaly detection [3], [4].

However, training ML models requires high-quality data that correctly represents the network activity of an organization. Despite containing valuable network traffic flows, the publicly available datasets also often include redundant features with noisy and missing data, which make the models less robust and slower [5], [6] . Therefore, to improve the detection performance and computational efficiency of the models, it is necessary to carefully analyze multiple datasets and select only the most relevant features for a cyber-attack detection task [7], [8].

Furthermore, the trained ML models must also perform well against adversarial examples, which are malicious traffic flows that contain specialized perturbations to be misclassified as benign [9]. Adversarial attacks can use these perturbations to deceive ML models, so it is also important to select the features that provide the best defense against such attacks [10]. By benchmarking the robustness of different ML models with different feature sets of different datasets, it is possible to identify the most suitable approaches for the computer networks of distinct organizations [11], [12].

This work presents the selection of the most relevant features of multiple network intrusion detection datasets, to be used to simultaneously improve the robustness and computational efficiency of ML models in the cybersecurity domain. Five feature selection methods, Information Gain, Chi-Squared Test, Recursive Feature Elimination, Mean Absolute Deviation, and Dispersion Ratio, were applied to the original CICIDS2017 dataset, a corrected version of it designated as NewCICIDS, the original HIKARI21 dataset, and an improved version of it designated as NewHIKARI.

For each dataset, two different feature sets were used to train ML models, one with only time-related characteristics and another with more specifically selected relevant features. Four types of ML models, Random Forest (RF), Extreme Gradient Boosting (XGB), Light Gradient Boosting Machine (LGBM), and Explainable Boosting Machine (EBM), were trained with regular training and adversarial training processes. Finally, the adversarial robustness benchmark carried out in [13] was extended to analyze the reliability of the different feature sets and their impact on the susceptibility of the models to adversarial examples of malicious network traffic flows.

The present paper is organized into multiple sections, meant to enable researchers to replicate this feature selection process for other datasets and perform trustworthy comparisons with the results of future studies. Section 2 provides a survey of previous work using feature selection methods for cybersecurity. Section 3 describes the datasets, the selection methods, the ML models, and the benchmark methodology. Section 4 presents the feature selection and consensus process to obtain the most relevant features. Section 5 presents an analysis of the obtained results in the benchmark. Lastly, Section 6 addresses the main conclusions and future research topics.

## 2   Related Work

To perform a reliable feature selection process and a trustworthy benchmark of ML models, it is important to understand the results and conclusions of previous work addressing network intrusion detection. Due to the complex tabular data structure of network traffic flows, where several features may be correlated, the presence of a given value at a given feature may restrict the values that other features can have [14]. Consequently, the search for optimal performance from ML models has been an ongoing quest, with challenges related to large quantities of noisy data, missing data, and lack of data diversity [15]. Over the last few years, several studies have used statistical methods to select the most relevant features and discard the redundant ones. The most relevant studies that use network traffic flows are described below.

A relatively recent study [16] demonstrated that a decision tree-based classification model can reach an accuracy of 80.60% and an F1-score of approximately 80% on the NSL-KDD dataset using only a reduced set of features carefully selected based on the information gain method and the chi-squared statistical test. In another study [17], an ensemble approach employing techniques such as information gain, gain ratio, chi-square and symmetric uncertainty were used to obtain a simplified set of 9 predictive features, and an RF was trained and obtained a very high accuracy of 98.90% on an Internet-of-Things (IoT) dataset for cyber-attack detection.

Another approach featured the use of statistical tools such as the Spearman rank correlation coefficient and the chi-squared statistic, together with a decision tree classifier [18]. This strategy led to a substantially reduced number of features, producing a F1-score result of 99.87% for the binary classification and 99.49% for the multi-class classification. Analyzing further into the field of feature selection, a two-phase methodology was proposed by [19], combining information gain with recursive feature elimination. This process resulted in the identification of 16 features from an IoT dataset, achieving a detection accuracy of 99.80% with an artificial neural network. Despite IoT network traffic being different from standard computer network traffic, an equivalent feature selection process can be used in enterprise networks.

With the aim of improving the learning performance of RF classifiers, researchers have explored new approaches [20], by integrating information gain to calculate the value of each feature, and the relief algorithm to calculate each feature weight, which resulted in improved ML model performance on the NSL-KDD dataset. Furthermore, in [21], recursive feature elimination was used together with an RF on the CICIDS2017 dataset, identifying the 4 most relevant features for a simple and efficient intrusion detection system. This approach resulted in an accuracy rate of 91% when applied to a deep neural network, but such a small number of features may leave the models more susceptible to small perturbations. Therefore, it is important to not reduce too much the number of features so that a model can achieve a good generalization while being robust against adversarial evasion attacks.

By applying the recursive feature elimination method on the NSL-KDD dataset, the study [22] selected the 13 best features, which were then used to train and evaluate different types of ML models, including RF, K-Nearest Neighbors and Naïve Bayes. This approach achieved an average accuracy of between 98% and 99%, although this approach was not validated on another dataset. Efforts were also made to find a balance between feature reduction and prediction accuracy in [23]. This study included cross-validation to ensure the reliability of the recursive feature elimination, and the size of the selected feature set was increased to 15. This improved the accuracy of the RF without increasing the required computational resources.

Computational efficiency is becoming a bigger concern, so multiple feature subsets of the CICIDS2017 dataset were analyzed in a study [24] that applied the information gain method. The results of this approach were applied to multiple ML models, and the highest accuracy was obtained by the J48 classifier with 52 features, a value of 99.87%. Nonetheless, an RF classifier reached 99.86% accuracy using only the 22 most relevant features. Therefore, ML models, and more specifically decision tree ensembles, can preserve a good generalization with a smaller feature set, which has the potential of reducing execution times and increasing both efficiency and robustness.

Overall, the recent studies have demonstrated the effectiveness of statistical feature selection methods to improve the performance of ML models for network intrusion detection. However, as new cyber-attacks and adversarial attacks are encountered, it is essential to analyze the features of more up-to-date datasets and how they impact the robustness of ML models against adversarial evasion attacks [25]. To the best of our knowledge, no previous work has analyzed how the time-related characteristics and more specifically selected features of the considered datasets affect the robustness of RF, XGB, LGBM, and EBM against adversarially perturbed network traffic flows.

## 3    Methods

This section describes the datasets, the selection methods, the ML models, and the benchmark methodology. The work was carried out on a machine with 16GB of RAM, a 6-core CPU and a 4GB GPU, which are reasonably common computation resources. The implementation relied on the Python programming language and the following libraries: *numpy* and *pandas* for general data manipulation, *xgboost* for the implementation of XGB, *lightgbm* for LGBM, *interpret* for EBM, and *scikit-learn* for the implementation of RF of the feature selection methods.

### 3.1    Datasets and Selection Methods

Due to their value for binary network traffic classification, four standard network traffic flow datasets were considered for the feature selection process: CICIDS2017, NewCICIDS, HIKARI21, and NewHIKARI. Their main characteristics are described below.

The CICIDS2017 [26] dataset is a very highly used dataset that contains common cyber-attacks performed in an enterprise computer network. It includes multiple captures of benign activity and several types of probing, brute-force, and DoS attacks, which were recorded in 2017 in an heterogenous testbed environment with 12 interacting machines. The network traffic flows were converted to a tabular data format using the CICFlowMeter [27] tool, provided by the Canadian Institute for Cybersecurity. The combined Tuesday and Wednesday traffic captures resulted in a total of 872105 data samples of the benign class and 266507 of the malicious class.

A corrected version of this dataset has been created to provide more realistic network traffic flows, which is designated as NewCICIDS [28], [29] in this work. Even though CICIDS2017 continues to be used as a standard benchmark dataset to compare the performance of novel ML models with baseline models from previous studies, some discrepancies have been noticed on a portion of the attack vectors it contains. The corrected version addressed this issue by correcting most of samples, although it has a reduced size, with 638432 benign and 106538 malicious samples.

The more recent HIKARI21 [30] dataset is starting to be used in various studies because it includes cyber-attacks that have started to be performed in more recent years. It contains probing and brute-force attacks, as well as benign background traffic of the normal operation of an enterprise computer network that uses the HTTPS communication protocol to encrypt network traffic. The data was recorded in 2021 to tackle the lack of datasets containing application-layer attacks on encrypted traffic, using similar features to those utilized in CICIDS2017. The resulting network flows correspond to 517582 benign samples and 37696 malicious samples.

An improved version of HIKARI21 has been released by its authors, which is designated as NewHIKARI [31] in this work. It contains a slightly lower number of benign samples, 214904, and almost a third of the malicious samples, 13349. Despite the reduced number of network traffic flows recorded in this dataset, the data samples represent more cyber-attack variations to include more recent cybersecurity threats. NewHIKARI has a higher class imbalance than the previous three datasets, representing more realistic conditions for enterprise-scale network intrusion detection.

The four datasets required a data preprocessing stage. In addition to creating stratified training and holdout sets with 70% and 30% of each dataset, it was necessary to select relevant and unbiased features that correctly represented the network activity. To obtain the feature importance rankings of the more than 80 features of each of these datasets and identify the most impactful ones, five methods were considered: Information Gain, Chi-Squared Test, Recursive Feature Elimination, Mean Absolute Deviation, and Dispersion Ratio. These selection methods are detailed below.

**Information Gain.** The concept of information gain is widely used in information theory to quantify the improvement in predictive ability [32]. It evaluates the reduction in uncertainty when a feature is included, according to the difference in entropy before and after considering that feature [33]. The mutual information method was utilized with the number of neighbors set to 3, to prevent the introduction of bias. For a feature $X$, the information gain $IG(X)$ can be obtained via the computation of the entropy in relation to the target class label:

$$IG(X) = H(Y) - H(Y|X)$$

where $H(Y)$ is the entropy of the target $Y$ and $H(Y|X)$ is the conditional entropy of the target given $X$.

**Chi-Squared Test.** The chi-squared statistical test is commonly used across the scientific community to assess the degree of dependence between a term and a class. It is fitted to the chi-squared continuous probability distribution of independent standard normal random variables, with one degree of freedom for analysis [34]. Considering a term $t$ and a class $c$, the test $CST(t, c)$ can be represented as:

$$CST(t,c) = \frac{S \times (PN - MQ)^2}{(P + M) \times (Q + N) \times (P + Q) \times (M + N)}$$

where $S$ is the total number of data samples, $P$ is the count of samples within class $c$ that contain $t$, $Q$ is the count of samples that contain the term $t$ but are not in class $c$, $M$ is the count of samples that belong to class $c$ but do not contain the term $t$, and $N$ is the count of samples from other classes without the term $t$.

**Recursive Feature Elimination.** The recursive approach to feature elimination [35] was originally a gene selection technique that required a classifier for the selection process. According to the weight assigned to each feature by the classifier, this method selects features by recursively considering smaller and smaller feature sets until one last feature remains. To obtain the ranking of each feature, the method was configured to only remove one feature per iteration of the recursive process.

**Mean Absolute Deviation.** The concept of mean absolute deviation is used as a scaling parameter within the Laplace distribution, providing a direct measure of the dispersion inherent in a feature. It is commonly employed as an alternative for the standard deviation, and when used as a feature selection method, it can identify and provide an ordered list of the features with the biggest impact to a classification task [36]. The deviation $MAD$ of a feature is mathematically defined as:

$$MAD = \frac{1}{n} \sum_{i=i}^{n} |X_i - \bar{X}|$$

where n is the total number of samples, $X_i$ is the value of sample number $i$, and $\bar{X}$ is the mean value of all samples.

**Dispersion Ratio.** The dispersion ratio of a feature is defined as the square root of the ratio of two components. The numerator represents the dispersion of the relative importance of a feature between the different classes, and the denominator represents the dispersion in the importance of that feature across the entire dataset [37]. This method obtains the relevance of a feature by calculating the ratio of the arithmetic mean to that of the geometric mean of a feature.

### 3.2   Models and Benchmark Methodology

The robustness analysis methodology introduced in [38] was followed to ensure an unbiased benchmark. It includes both a regular training process and an adversarial training process, which is a well-established adversarial defense strategy. In the former, the original training set of a certain dataset is used to train, fine-tune, and validate an ML model. In the latter, data augmentation is performed by creating simple perturbations in the original training set, resulting in an adversarial training set that contains both original data samples and slightly perturbed data samples.

Afterwards, the considered methodology establishes a performance evaluation in both normal conditions and during a direct attack to the models. In the former, the models perform predictions of the data samples in the regular holdout set of a certain dataset, and several standard evaluation metrics are computed. In the latter, a full adversarial evasion attack is performed against each model, with specialized perturbations to deceive that specific model. Since different models are susceptible to different perturbations, the attacks result in model-specific adversarial holdout sets. In the case of network intrusion detection, these attacks are targeted, attempting to cause misclassifications from the malicious class to the target benign class.

The adversarial examples were generated using the Adaptative Perturbation Pattern Method (A2PM) [39]. It relies on pattern sequences that learn the characteristics of each class and create constrained data perturbations, according to the provided information about the feature set, which corresponds to a gray-box setting. The patterns record the value intervals of different feature subsets, which are then used to ensure that the perturbations take the correlations of the features into account, generating realistic adversarial examples. Therefore, when applied to network intrusion detection, the patterns iteratively optimize the perturbations that are performed on each feature of a network traffic flow according to the constraints of a computer network.

For adversarial training, a simple function provided by A2PM was used to create a different perturbation in one of the features of each malicious sample of a training set, performing data augmentation to increase data diversity. Hence, a model was able to learn not only from a cyber-attack flow, but also from a simple variation of that flow. Starting from a regular training set with 70% of a dataset, another set of the same size can be obtained, with a perturbation in each sample.

To perform adversarial evasion attacks specific to each model, the full adversarial attack created as many data perturbations as necessary in a holdout set until every malicious sample was misclassified as benign or a total of 15 attack iterations were performed. No more iterations were allowed because a high number of requests to a specific server would increase the risk of the anomalous behavior being noticed by the security practitioners overseeing the networking infrastructure of an enterprise network [40]. Starting from a regular holdout set with 30% of a dataset, several other sets of the same size can be obtained, with specialized perturbations for a specific model.

Due to their well-established performance in network intrusion detection, four types of ML models were considered: RF, XGB, LGBM, and EBM. The optimal configuration for each model and each dataset were obtained via a grid search of well-established hyperparameter combinations, and the best ones were determined through a 5-fold cross-validation. The F1-score, which consolidates precision and recall and is suitable for imbalanced data, was selected as the validation metric. After the fine-tuning process, each model was retrained with a complete training set to be ready for the benchmark with the regular and adversarial holdout sets. The models and their optimized hyperparameters are described below.

**Random Forest.** RF [41] is a supervised ensemble created through bagging and using the Gini Impurity criterion to calculate the best node splits. Each individual tree performs a prediction according to a feature subset, and the most common vote is chosen. RF is based on the concept that the collective decisions of many trees will be better than the decisions of just one. Table 1 summarizes the fine-tuned configuration.

**Table 1.** Summary of RF configuration.

| Parameter | Value |
| --- | --- |
| Criterion | Gini Impurity |
| No. of estimators | 100 |
| Max. features | 4 |
| Max. depth of a tree | 8 to 16 |
| Min. samples in a leaf | 1 to 4 |

**Extreme Gradient Boosting.** XGB [42] performs gradient boosting using a supervised ensemble with a level-wise growth strategy. The nodes within each tree are split level by level, using the Histogram method to compute fast histogram-based approximations and seeking to minimize the Cross-Entropy loss function during its training. Table 2 summarizes the fine-tuned configuration.

**Table 2.** Summary of XGB configuration.

| Parameter | Value |
| --- | --- |
| Method | Histogram |
| Loss function | Cross-Entropy |
| No. of estimators | 100 |
| Max. depth of a tree | 4 to 16 |
| Min. leaf weight | 1 |
| Min. loss reduction | 0.01 |
| Learning rate | 0.1 to 0.3 |
| Feature subsample | 0.7 to 0.9 |

**Light Gradient Boosting Machine.** LGBM [43] also uses a supervised ensemble to perform gradient boosting. The nodes are split using a leaf-wise strategy for a best-first approach, performing the split with the higher loss reduction. LGBM uses Gradient-based One-Side Sampling (GOSS) to build the decision trees, which is computationally lighter than previous methods and therefore provides a faster training process. Table 3 summarizes the fine-tuned configuration.

**Table 3.** Summary of LGBM configuration.

| Parameter | Value |
| --- | --- |
| Method | GOSS |
| Loss function | Cross-Entropy |
| No. of estimators | 100 |
| Max. leaves in a tree | 15 |
| Min. samples in a leaf | 20 |
| Min. loss reduction | 0.01 |
| Learning rate | 0.1 to 0.2 |
| Feature subsample | 0.7 to 0.8 |

**Explainable Boosting Machine.** EBM [44] is a generalized additive model that performs cyclic gradient boosting with a tree ensemble. Unlike the other three black-box models, EBM is a glass-box model that remains explainable and interpretable during the inference phase [45]. Each feature contributes to a prediction in an additive manner that enables their individual contribution to be measured and explained. Table 4 summarizes the fine-tuned configuration.

Table 4. Summary of EBM configuration.

| Parameter | Value |
|---|---|
| Loss function | Cross-Entropy |
| No. of estimators | 100 |
| Max. number of bins | 256 |
| Max. leaves in a tree | 3 to 15 |
| Min. samples in a leaf | 1 to 2 |
| Learning rate | 0.1 |

## 4 Feature Selection

This section presents the obtained feature sets and highlights the most relevant features. Initially, the considered datasets contained 83 features, which were similar in both CICIDS2017, HIKARI, and their novel versions. The five feature selection methods were applied independently to each dataset, and since the results provided by the different methods were in different orders of magnitude, they were normalized to the range of zero to one. Therefore, the output of each method was re-scaled to assign a percentage of relevance to each feature, with the sum of all features being 100%.

A consensus process was designed with a minimum relevance threshold of 1% to only select features that contribute to the detection and disregard those that at least one method considers to be negligible. The results of each method were analyzed independently, and the features with a relevance value below the threshold were excluded. Then, the five methods were combined, and a ranking was obtained by computing the mean relevance value of each remaining feature. A common feature set was created, consisting of 24 time-related features present in all the considered datasets. Then, the consensus process was applied to each dataset, resulting in 26 features specifically for CICIDS2017 and NewCICIDS, and 22 for HIKARI21 and NewHIKARI.

The first feature set, containing the 24 common features, considered 7 main time-related characteristics of a network traffic flow. In the considered datasets, the forward part of a flow corresponds to a client machine that opens a connection with the server, sending network packets. Likewise, the backward part corresponds to the packets sent by the server back to the client within that connection. The full connection will be classified as either a benign flow that is part of the normal operation of the network or a malicious flow in which the client sent ill-intentioned packets. Regarding the IAT keyword, it corresponds to the Inter-Arrival Time, the elapsed time between the arrival of two subsequent network packets within a flow. Table 5 provides an overview of the selected time-related characteristics.

**Table 5.** Selected time-related characteristics.

| Characteristic | Description | Selected Features | | | | |
|---|---|---|---|---|---|---|
| | | Total | Mean | Std | Max | Min |
| Flow Packet IAT | Packet IAT of the full connection | No | Yes | Yes | Yes | Yes |
| Forward Packet IAT | Packet IAT of the client | Yes | Yes | Yes | Yes | Yes |
| Backward Packet IAT | Packet IAT of the server | Yes | Yes | Yes | Yes | Yes |
| Forward Bulk Rate | Transmission rate of the client | No | Yes | No | No | No |
| Backward Bulk Rate | Transmission rate of the server | No | Yes | No | No | No |
| Flow Active Time | Transmission time of the full connection | No | Yes | Yes | Yes | Yes |
| Flow Idle Time | Inactive time of the full connection | No | Yes | Yes | Yes | Yes |

Regarding the specific feature set for CICIDS2017, from all the considered features, a total of 26 exceeded the 1% threshold. Features related to the idle time appeared in the top 14, representing collectively a significant 24% of the total relevance, providing a distinct view of the dataset. Furthermore, the characteristics relating to the IAT showed a notable relevance, each surpassing the 1% threshold and contributing nearly 26%. The Maximum and Mean Idle Time also revealed their major roles as the two most relevant features, together representing 16% of the total relevance. Table 6 provides the ranking for CICIDS2017, with Fwd corresponding to Forward, Bwd to Backward, Max to Maximum, Min to Minimum, and Std to Standard Deviation.

**Table 6.** Feature ranking for CICIDS2017.

| No. | Feature | No. | Feature | No. | Feature |
|---|---|---|---|---|---|
| 1 | Max. Idle Time | 10 | Fwd. Max. IAT | 19 | Mean Payload Size |
| 2 | Mean Idle Time | 11 | Bwd. Total IAT | 20 | Bwd. Std. IAT |
| 3 | Flow Duration | 12 | Bwd. Min. IAT | 21 | Bwd. Mean Segment Size |

| No. | Feature | No. | Feature | No. | Feature |
|---|---|---|---|---|---|
| 4 | Min. Idle Time | 13 | Max. IAT | 22 | Min. IAT |
| 5 | Connection Flag | 14 | Std. Idle Time | 23 | Fwd. Mean IAT |
| 6 | Fwd. Header Size | 15 | Bwd. Max. IAT | 24 | Std. IAT |
| 7 | Fwd. Min. IAT | 16 | Destination Port | 25 | Bwd. Mean IAT |
| 8 | Fwd. Total IAT | 17 | Bytes Per Second | 26 | Mean IAT |
| 9 | Fwd. Std. IAT | 18 | Std. Payload Size | | |

Regarding NewCICIDS, among the 26 features that went beyond the 1% threshold, there were several interesting results. Apart from the minimum IAT, all the features related to IAT exhibited a significance of more than 1%, adding up to a combined score of over 37%. All the features related to Idle Time got significant consideration by most methods, ranking in the top 10 and for 20% of the total relevance. The features associated to the duration and IAT at origin and destination and vice versa were found to be highly important, occupying the top 3 positions with a combined relevance of over 16%. Table 7 provides the ranking for NewCICIDS.

**Table 7.** Feature ranking for NewCICIDS.

| No. | Feature | No. | Feature | No. | Feature |
|---|---|---|---|---|---|
| 1 | Bwd. Total IAT | 10 | Std. Idle Time | 19 | Fwd. Std. IAT |
| 2 | Flow Duration | 11 | Bwd. Max. IAT | 20 | Mean IAT |
| 3 | Fwd. Total IAT | 12 | Bwd. Min. IAT | 21 | Communication protocol |
| 4 | Std. Payload Size | 13 | Fwd. Max. IAT | 22 | Bwd. Mean Payload Size |
| 5 | Max. Idle Time | 14 | Max. IAT | 23 | Bwd. Max. Payload Size |
| 6 | Mean Idle Time | 15 | Mean Payload Size | 24 | Bwd. Total Subflow |
| 7 | Connection Flags | 16 | Fwd. Mean IAT | 25 | Bwd. Std. Payload Size |
| 8 | Min. Idle Time | 17 | Bwd. Mean IAT | 26 | Bwd. Std. IAT |
| 9 | Fwd. Min. IAT | 18 | Std. IAT | | |

Regarding the specific feature set for HIKARI21, less than a third of all the considered features managed to achieve a value of calculated relevance of 1%. From the final 22 selected characteristics, several notable features stood out. Most of the characteristics related to IAT exceeded the designated threshold, and these together contributed 23% of the total relevance. Other features related to Idle Time showed a higher relevance, with all of them ranking in the top 12, a ranking in the top 5. These characteristics score together accounted for more than 26% of the total relevance. Backward Bulk Rate achieved the first place, demonstrating an importance one and a half times superior to that of the feature in second place and three times that of the feature in third place. The combined relevance score of the top 3 features reached nearly 31%. Table 8 provides the ranking for HIKARI21.

**Table 8.** Feature ranking for HIKARI21.

| No. | Feature | No. | Feature | No. | Feature |
|---|---|---|---|---|---|
| 1 | Backward Bulk Rate | 9 | Connection Flags | 17 | Mean Active Time |
| 2 | Total Idle Time | 10 | Fwd. Max. IAT | 18 | Fwd. Mean IAT |
| 3 | Max. Idle Time | 11 | Max. IAT | 19 | Fwd. Min. IAT |

| No. | Feature | No. | Feature | No. | Feature |
|---|---|---|---|---|---|
| 4 | Mean Idle Time | 12 | Std. Idle Time | 20 | Min. IAT |
| 5 | Min. Idle Time | 13 | Bwd. Max. IAT | 21 | Fwd. Std. IAT |
| 6 | Total IAT | 14 | Min. Active Time | 22 | Mean IAT |
| 7 | Fwd. Total IAT | 15 | Max. Active Time | | |
| 8 | Bwd. Total IAT | 16 | Total Active Time | | |

Regarding NewHIKARI, 22 managed to obtain relevance values of more than 1%, making up less than a third of the total features initially evaluated. Of this final selection, some characteristics turned out to be notable, such as all the Idle Time features, that ranked in the top 6, apart from the Standard Deviation of the Active Time, which fell below the threshold. Together, these features accounted for 27% of the total relevance. Two thirds of the IAT related features met the 1% threshold, collectively contributing to 25%. The characteristic with the highest ranking, Total Idle Time, was nearly twice as important as the second most significant feature in the dataset, while the following ranking from 2 to 6 showed similar values, between 5% and 6%. Table 9 provides the ranking for NewHIKARI.

**Table 9.** Feature ranking for NewHIKARI.

| No. | Feature | No. | Feature | No. | Feature |
|---|---|---|---|---|---|
| 1 | Total Idle Time | 9 | Total Active Time | 17 | Min. IAT |
| 2 | Fwd. Total IAT | 10 | Mean Active Time | 18 | Mean IAT |
| 3 | Max. Idle Time | 11 | Min. Active Time | 19 | Fwd. Min. IAT |
| 4 | Mean Idle Time | 12 | Bytes Per Second | 20 | Fwd. Mean IAT |
| 5 | Total IAT | 13 | Max. IAT | 21 | Total Payload Size |
| 6 | Min. Idle Time | 14 | Fwd. Max. IAT | 22 | Fwd. Total Payload Size |
| 7 | Max. Active Time | 15 | Bwd. Total IAT | | |
| 8 | Connection Flag | 16 | Bwd. Max. IAT | | |

## 5   Results and Discussion

This section presents and discusses the results obtained by evaluating the performance of the ML models created with the two different sets, as well as with regular and adversarial training. The evaluation considers the regular holdout set of the CICIDS2017, NewCICIDS, HIKARI21, and NewHIKARI datasets, and the model-specific adversarial holdout sets created by the adversarial attacks.

The benchmark considered standard evaluation metrics for binary network traffic classification. The ACC, PRC, RCL, F1S, and FPR columns of the following tables correspond to accuracy, precision, recall, F1-score, and false positive rate. The optimal result would be 100% for all metrics except the false positive rate, which should be as close to 0% as possible. Additionally, the results achieved by the adversarially trained models on the adversarial holdout sets are highlighted in bold in all tables.

### 5.1 CICIDS2017

The models trained with the CICIDS2017 dataset obtained very good results across all the evaluation metrics. The time-related feature set enabled all four models to detect the anomalous behavior of most malicious flows, distinguishing cyber-attacks from benign activity and reaching F1-scores over 89%. Nonetheless, when adversarial attacks were performed against these models, their precision and recall exhibited significant declines that resulted in F1-scores lower than 0.1% after the attack iterations were complete. This failure to detect adversarial examples suggests that ensembles of decision trees are inherently vulnerable to modifications of the time-related characteristics of network traffic flows.

On the other hand, the models created through adversarial training had substantially lower declines, preserving their precision above 97%. Even though the recall of EBM was only approximately 60% when attacked, RF, XGB, and LGBM all retained a higher recall above 73%. Hence, by training with slightly perturbed malicious samples, the robustness of the models was improved, and most malicious flows could not evade detection. Regarding benign flows, it is important to note that the false positive rates were decreased to below 0.40%, which indicates that deploying these models in a real computer network could lead to less false alarms. Table 10 provides the results of the models trained with the time-related feature set.

Table 10. Obtained results for CICIDS2017 with time-related features.

| Model | Training | Attacked | Evaluation Metrics (%) | | | | |
|---|---|---|---|---|---|---|---|
| | | | ACC | PRC | RCL | F1S | FPR |
| RF | Regular | No | 95.21 | 90.74 | 88.59 | 89.65 | 2.76 |
| | | Yes | 74.48 | 0.03 | 0.01 | 0.01 | 2.76 |
| | Adversarial | No | 93.84 | 99.20 | 74.30 | 84.96 | 0.18 |
| | | Yes | **93.80** | **99.19** | **74.12** | **84.84** | **0.18** |
| XGB | Regular | No | 95.29 | 90.73 | 88.94 | 89.83 | 2.78 |
| | | Yes | 74.48 | 0.36 | 0.03 | 0.06 | 2.78 |
| | Adversarial | No | 94.67 | 98.93 | 78.07 | 87.27 | 0.26 |
| | | Yes | **94.30** | **98.91** | **76.5** | **86.28** | **0.26** |
| LGBM | Regular | No | 94.95 | 90.12 | 88.09 | 89.09 | 2.95 |
| | | Yes | 74.35 | 0.57 | 0.05 | 0.09 | 2.95 |
| | Adversarial | No | 94.19 | 98.93 | 76.01 | 85.97 | 0.25 |
| | | Yes | **93.50** | **98.89** | **73.05** | **84.03** | **0.25** |
| EBM | Regular | No | 94.98 | 90.43 | 87.86 | 89.13 | 2.84 |
| | | Yes | 74.42 | 0.08 | 0.01 | 0.01 | 2.84 |
| | Adversarial | No | 94.4 | 98.43 | 77.31 | 86.6 | 0.38 |
| | | Yes | **90.13** | **97.96** | **60.08** | **74.48** | **0.38** |

The feature set obtained through the combination of multiple feature selection methods led to substantial improvements in all four models. Their accuracy was approximately 3% higher, and their improved precision and recall led to F1-scores 8% higher. This difference was even higher when the models were attacked, which led to 12% higher F1-scores in RF, XGB, and LGBM, and even 23% higher in EBM. Since the benchmark was performed in the same conditions and with fine-tuned models, these results demonstrate the impact that different feature sets can have on the robustness of ML models for a cyber-attack detection task.

It is pertinent to highlight that the false positive rate of the regularly trained models was lowered to less than half. Nonetheless, it could not be further reduced with adversarial training, remaining higher than with the previous time-related features. This suggests that, despite the benefits of the more specific features, they will lead to a greater quantity of false alarms, which would cause an organization to spend resources and time on unnecessary mitigation measures. Table 11 provides the results of the models trained with the more specific feature set.

Table 11. Obtained results for CICIDS2017 with feature selection.

| Model | Training | Attacked | Evaluation Metrics (%) | | | | |
|---|---|---|---|---|---|---|---|
| | | | ACC | PRC | RCL | F1S | FPR |
| RF | Regular | No | 98.84 | 96.84 | 98.27 | 97.55 | 0.98 |
| | | Yes | 86.56 | 93.45 | 45.79 | 61.46 | 0.98 |
| | Adversarial | No | 98.58 | 97.46 | 96.45 | 96.95 | 0.77 |
| | | Yes | **98.58** | **97.46** | **96.44** | **96.95** | **0.77** |
| XGB | Regular | No | 98.86 | 98.86 | 98.30 | 97.58 | 0.97 |
| | | Yes | 77.76 | 71.90 | 8.15 | 14.65 | 0.97 |
| | Adversarial | No | 98.80 | 97.26 | 97.60 | 97.43 | 0.84 |
| | | Yes | **98.79** | **97.26** | **97.60** | **97.43** | **0.84** |
| LGBM | Regular | No | 98.80 | 96.68 | 98.26 | 97.46 | 1.03 |
| | | Yes | 83.59 | 90.78 | 33.25 | 48.68 | 1.03 |
| | Adversarial | No | 98.67 | 97.22 | 97.11 | 97.16 | 0.85 |
| | | Yes | **98.67** | **97.22** | **97.09** | **97.15** | **0.85** |
| EBM | Regular | No | 98.68 | 96.44 | 97.95 | 97.19 | 1.10 |
| | | Yes | 78.66 | 77.49 | 12.43 | 21.43 | 1.10 |
| | Adversarial | No | 98.63 | 97.01 | 97.14 | 97.07 | 0.92 |
| | | Yes | **98.57** | **96.99** | **96.90** | **96.95** | **0.92** |

### 5.2  NewCICIDS

The corrected version of CICIDS2017 exhibited better results than those of the original dataset, using the time-related features. Training with the corrected network traffic flows of NewCICIDS led all four models to achieve F1-scores higher than 99% on the regular holdout set, and their false positive rates did not exceed 0.20%. Despite the performance of the models being significantly decreased in the model-specific adversarial holdout sets, it was still slightly better than the decline observed in the original dataset. It is important to note the better robustness of the regularly trained EBM, which retained a precision of over 31% throughout the adversarial evasion attack just by training with the corrected flows of NewCICIDS.

As before, performing adversarial training led to a great improvement in the robustness of the models. This defense strategy enabled the detection of most adversarial cyber-attack examples, reducing the number of misclassifications that would be harmful for an enterprise. Even though the recall of the adversarially trained RF and XGB was slightly decreased, they preserved their precision of 99.88% and 99.90%, without this metric being decreased by the attack. Since only very few of the perturbed malicious flows were misclassified, the functionality of those cyber-attacks would be prevented in a real enterprise communication network. Furthermore, RF and XGB achieved the best false positive rates, 0.06% and 0.05%, respectively, which indicates that the corrected dataset also led to models with better generalization. Table 12 provides the results of the time-related feature set, highlighting the good precision retained by the regularly trained EBM.

**Table 12.** Obtained results for NewCICIDS with time-related features.

| Model | Training | Attacked | Evaluation Metrics (%) | | | | |
|---|---|---|---|---|---|---|---|
| | | | ACC | PRC | RCL | F1S | FPR |
| RF | Regular | No | 99.90 | 99.81 | 99.92 | 99.87 | 0.11 |
| | | Yes | 64.19 | 0.01 | 0.01 | 0.01 | 0.11 |
| | Adversarial | No | 99.67 | 99.88 | 99.20 | 99.54 | 0.06 |
| | | Yes | **99.63** | **99.88** | **99.08** | **99.48** | **0.06** |
| XGB | Regular | No | 99.94 | 99.89 | 99.94 | 99.92 | 0.06 |
| | | Yes | 64.22 | 2.59 | 0.01 | 0.01 | 0.06 |
| | Adversarial | No | 99.93 | 99.90 | 99.89 | 99.90 | 0.05 |
| | | Yes | **99.84** | **99.90** | **99.64** | **99.77** | **0.05** |
| LGBM | Regular | No | 99.79 | 99.63 | 99.79 | 99.71 | 0.20 |
| | | Yes | 64.14 | 7.36 | 0.03 | 0.06 | 0.20 |
| | Adversarial | No | 99.72 | 99.67 | 99.54 | 99.60 | 0.19 |
| | | Yes | **94.91** | **99.61** | **86.10** | **92.36** | **0.19** |
| EBM | Regular | No | 99.86 | 99.76 | 99.84 | 99.80 | 0.14 |
| | | Yes | 64.21 | 31.48 | 0.11 | 0.22 | 0.14 |
| | Adversarial | No | 99.80 | 99.74 | 99.71 | 99.72 | 0.15 |
| | | Yes | **98.13** | **99.72** | **95.02** | **97.32** | **0.15** |

In NewCICIDS, the more specific features also led to some improvements in the results of XGB, LGBM, and EBM, across all the evaluation metrics considered in this benchmark. Both the regularly trained and the adversarially trained models exhibited a seemingly better robustness against the perturbations. The exception was RF, which obtained a final accuracy of 96.91% and a lower F1-score of 95.48% when attacked, even after the improvements of adversarial training. Despite still being a very high score, it was lower than the value obtained with the time-related features.

Nonetheless, the almost optimal results obtained by both feature sets of this dataset create some doubt about the realism of the data samples it contains. The performed corrections significantly improve the results of most ML models, but those results may have been reached only because some cyber-attack variations were removed in NewCICIDS, reducing its data diversity. Therefore, this dataset should be used with caution because an ML model must be trained with malicious flows that truly represent the cyber-attacks targeting a real computer network of a modern organization. Table 13 provides the results of the models trained with the more specific feature set.

Table 13. Obtained results for NewCICIDS with feature selection.

| Model | Training | Attacked | Evaluation Metrics (%) | | | | |
|---|---|---|---|---|---|---|---|
| | | | ACC | PRC | RCL | F1S | FPR |
| RF | Regular | No | 99.98 | 99.95 | 99.99 | 99.97 | 0.03 |
| | | Yes | 64.42 | 90.89 | 0.52 | 1.02 | 0.03 |
| | Adversarial | No | 99.97 | 99.95 | 99.97 | 99.96 | 0.03 |
| | | Yes | **96.91** | **99.95** | **91.40** | **95.48** | **0.03** |
| XGB | Regular | No | 99.98 | 99.96 | 99.99 | 99.98 | 0.02 |
| | | Yes | 66.19 | 99.23 | 5.45 | 10.34 | 0.02 |
| | Adversarial | No | 99.97 | 99.96 | 99.97 | 99.96 | 0.02 |
| | | Yes | **99.73** | **99.96** | **99.29** | **99.62** | **0.02** |
| LGBM | Regular | No | 99.97 | 99.93 | 99.98 | 99.96 | 0.04 |
| | | Yes | 68.90 | 99.48 | 13.05 | 23.07 | 0.04 |
| | Adversarial | No | 99.96 | 99.93 | 99.97 | 99.95 | 0.04 |
| | | Yes | **99.79** | **99.93** | **99.49** | **99.71** | **0.04** |
| EBM | Regular | No | 99.98 | 99.95 | 99.99 | 99.97 | 0.03 |
| | | Yes | 67.45 | 99.41 | 8.98 | 16.47 | 0.03 |
| | Adversarial | No | 99.98 | 99.94 | 99.99 | 99.97 | 0.03 |
| | | Yes | **99.65** | **99.94** | **99.07** | **99.50** | **0.03** |

### 5.3 HIKARI21

The HIKARI21 dataset with more up-to-date network traffic flows enabled the ML models to obtain a relatively similar accuracy to CICIDS2017 with the time-related features, approximately 93%. However, this metric does not express the impact of the class imbalance between benign and malicious traffic flows. It can be observed that even though the models correctly classified most benign flows, a very low recall was obtained because the malicious flows could not be detected. This led to F1-scores under 33% in the regular holdout sets, and scores under 1% when RF, XGB, and LGBM were attacked. Despite EBM being able to retain 3.18%, these results demonstrate that using only the time-related features was not an adequate approach for this dataset.

Even when adversarial training was performed, none of the four ML models was able to detect the adversarial examples, which resulted in a very large number of misclassifications and very low F1-scores. These scores are substantially lower than those obtained in the previous datasets, suggesting that the greater complexity of the more recent cyber-attacks makes it more difficult to distinguish them from benign flows that are part of the normal operation of an enterprise computer network. Table 14 provides the results of the models trained with the time-related feature set.

**Table 14.** Obtained results for HIKARI21 with time-related features.

| Model | Training | Attacked | Evaluation Metrics (%) | | | | |
|---|---|---|---|---|---|---|---|
| | | | ACC | PRC | RCL | F1S | FPR |
| RF | Regular | No | 92.01 | 30.96 | 14.42 | 19.68 | 2.34 |
| | | Yes | 91.03 | 0.01 | 0.01 | 0.01 | 2.34 |
| | Adversarial | No | 93.20 | 13.33 | 0.02 | 0.04 | 0.01 |
| | | Yes | **93.20** | **7.14** | **0.01** | **0.02** | **0.01** |
| XGB | Regular | No | 93.16 | 49.24 | 23.58 | 31.89 | 1.77 |
| | | Yes | 91.56 | 0.01 | 0.01 | 0.01 | 1.77 |
| | Adversarial | No | 93.20 | 36.73 | 0.16 | 0.32 | 0.02 |
| | | Yes | **93.20** | **35.42** | **0.15** | **0.30** | **0.02** |
| LGBM | Regular | No | 93.25 | 50.66 | 23.39 | 32.01 | 1.66 |
| | | Yes | 91.66 | 0.01 | 0.01 | 0.01 | 1.66 |
| | Adversarial | No | 93.20 | 16.13 | 0.04 | 0.09 | 0.02 |
| | | Yes | **93.20** | **10.24** | **0.03** | **0.05** | **0.02** |
| EBM | Regular | No | 93.17 | 49.37 | 24.41 | 32.67 | 1.82 |
| | | Yes | 91.51 | 0.01 | 0.01 | 0.01 | 1.82 |
| | Adversarial | No | 93.11 | 39.83 | 2.94 | 5.48 | 0.32 |
| | | Yes | **93.02** | **27.52** | **1.69** | **3.18** | **0.32** |

The feature set obtained through the combination of multiple feature selection methods did not provide substantial improvements. The results were similar to those of the time-related features in both evaluations, with regular malicious flows and with adversarial examples of those flows. This demonstrates that even when the best possible features were selected for this dataset, neither RF, XGB, LGBM, nor EBM could obtain better results than when only the time-related characteristics of the flows were considered.

Furthermore, adversarial training is not always guaranteed to help ML models achieve an adversarially robust generalization. When the models cannot achieve good results even in the regular holdout set, adversarial training may not be the best approach, so other adversarial defense strategies should be explored. Table 15 provides the results of the models trained with the more specific feature set.

**Table 15.** Obtained results for HIKARI21 with feature selection.

| Model | Training | Attacked | Evaluation Metrics (%) | | | | |
|---|---|---|---|---|---|---|---|
| | | | ACC | PRC | RCL | F1S | FPR |
| RF | Regular | No | 92.22 | 33.27 | 14.47 | 20.16 | 2.11 |
| | | Yes | 91.24 | 0.01 | 0.01 | 0.01 | 2.11 |
| | Adversarial | No | 93.20 | 10.02 | 0.02 | 0.04 | 0.01 |
| | | Yes | **93.20** | **9.37** | **0.01** | **0.02** | **0.01** |
| XGB | Regular | No | 93.21 | 50.01 | 21.10 | 32.53 | 1.76 |
| | | Yes | 91.57 | 0.01 | 0.01 | 0.01 | 1.76 |
| | Adversarial | No | 93.21 | 33.34 | 0.07 | 0.14 | 0.01 |
| | | Yes | **93.21** | **30.43** | **0.06** | **0.12** | **0.01** |

| Model | Training | Attacked | ACC | PRC | RCL | F1S | FPR |
|---|---|---|---|---|---|---|---|
| LGBM | Regular | No | 93.26 | 50.82 | 21.98 | 30.69 | 1.55 |
| | | Yes | 91.77 | 0.01 | 0.01 | 0.01 | 1.55 |
| | Adversarial | No | 93.20 | 27.27 | 0.11 | 0.21 | 0.02 |
| | | Yes | **93.20** | **26.83** | **0.10** | **0.19** | **0.02** |
| EBM | Regular | No | 93.21 | 49.97 | 24.78 | 33.13 | 1.81 |
| | | Yes | 91.53 | 0.01 | 0.01 | 0.01 | 1.81 |
| | Adversarial | No | 93.18 | 43.48 | 1.68 | 3.24 | 0.16 |
| | | Yes | **93.17** | **42.56** | **1.62** | **3.12** | **0.16** |

### 5.4 NewHIKARI

The models trained with the improved version of the HIKARI21 dataset with more cyber-attack variations obtained significantly better results. Their F1-scores were between 83% and 84%, which can be valuable for network traffic analysis. The adversarial evasion attack caused the recall and precision of all four models to decrease, but XGB was able to retain a precision of over 58% and EBM of over 98%. Since their false positive rates were near 0.01%, the results denote that more than half of the adversarial examples were detected and there were very few benign flows mistakenly predicted as malicious, which is important for an enterprise-scale computer network.

Despite also having equivalent false positive rates, the adversarially trained models could not retain a high robustness. When attacked, the F1-scores of RF, EBM, XGB, and LGBM, decreased to 82%, 63%, 62%, and 27%. Even though RF remained close to its original score of 83%, it was observed that the more complex ensembles were much more vulnerable to the perturbations created by an adversarial attack. Even though these models generally have better results, it is pertinent to always evaluate them with different feature sets, seeking a good generalization to regular traffic and a good robustness to adversarially perturbed traffic. Table 16 provides the results of the time-related feature set, highlighting the precision of the regularly trained XGB and EBM.

**Table 16.** Obtained results for NewHIKARI with time-related features.

| Model | Training | Attacked | Evaluation Metrics (%) | | | | |
|---|---|---|---|---|---|---|---|
| | | | ACC | PRC | RCL | F1S | FPR |
| RF | Regular | No | 98.34 | 99.79 | 71.84 | 83.54 | 0.01 |
| | | Yes | 94.14 | 0.01 | 0.01 | 0.01 | 0.01 |
| | Adversarial | No | 98.33 | 99.90 | 71.59 | 83.40 | 0.01 |
| | | Yes | **98.21** | **99.89** | **69.46** | **81.94** | **0.01** |
| XGB | Regular | No | 98.35 | 99.83 | 71.91 | 83.60 | 0.01 |
| | | Yes | 94.15 | **58.33** | 0.17 | 0.35 | 0.01 |
| | Adversarial | No | 98.36 | 99.86 | 72.01 | 83.68 | 0.01 |
| | | Yes | **96.76** | **99.78** | **44.72** | **61.76** | **0.01** |
| LGBM | Regular | No | 98.36 | 99.86 | 72.13 | 83.76 | 0.01 |
| | | Yes | 94.15 | 0.01 | 0.01 | 0.01 | 0.01 |
| | Adversarial | No | 98.35 | 99.72 | 72.01 | 83.63 | 0.01 |
| | | Yes | **95.05** | **98.74** | **15.63** | **26.99** | **0.01** |
| EBM | Regular | No | 98.35 | 99.76 | 72.01 | 83.64 | 0.01 |
| | | Yes | 95.01 | **98.84** | 14.86 | 25.83 | 0.01 |
| | Adversarial | No | 98.35 | 99.69 | 71.94 | 83.57 | 0.01 |
| | | Yes | **96.84** | **99.52** | **46.14** | **63.05** | **0.01** |

In NewHIKARI, training the models with the more specific features also improved all the evaluation metrics. The adversarially trained XGB and EBM retained F1-scores of 84.76% and 83.99% when attacked, which are significantly higher than with the previous time-related features. LGBM stands out for reaching the highest F1-score when attacked, a value of 84.96%, which contrasts with the lowest score of 27% obtained before. Therefore, by using an improved dataset with more data diversity, selecting a more specific feature set, and combining it with adversarial training, the ML models were able to achieve a more robust generalization.

Since the false positive rates of all four ML models also remained very low, the employed feature selection process provided a good balance between true positives and false positives, which prevents the disruptions caused by the common cyber-attacks and their adversarial examples, and also minimizes the number of false alarms and unnecessary mitigation measures that would be costly for an organization. Table 17 provides the results of the models trained with the more specific feature set.

**Table 17.** Obtained results for NewHIKARI with feature selection.

| Model | Training | Attacked | Evaluation Metrics (%) | | | | |
|---|---|---|---|---|---|---|---|
| | | | ACC | PRC | RCL | F1S | FPR |
| RF | Regular | No | 98.47 | 99.90 | 73.86 | 84.93 | 0.01 |
| | | Yes | 94.47 | 98.65 | 5.49 | 10.41 | 0.01 |
| | Adversarial | No | 98.45 | 99.90 | 73.56 | 84.73 | 0.01 |
| | | Yes | **98.45** | **99.90** | **73.56** | **84.73** | **0.01** |
| XGB | Regular | No | 98.45 | 99.73 | 73.76 | 84.80 | 0.01 |
| | | Yes | 97.14 | 99.61 | 51.36 | 67.78 | 0.01 |
| | Adversarial | No | 98.45 | 99.76 | 73.73 | 84.80 | 0.01 |
| | | Yes | **98.45** | **99.76** | **73.68** | **84.76** | **0.01** |
| LGBM | Regular | No | 98.47 | 99.90 | 73.96 | 84.99 | 0.01 |
| | | Yes | 97.04 | 99.85 | 49.39 | 66.09 | 0.01 |
| | Adversarial | No | 98.47 | 99.90 | 73.93 | 84.98 | 0.01 |
| | | Yes | **98.47** | **99.90** | **73.91** | **84.96** | **0.01** |
| EBM | Regular | No | 98.45 | 99.66 | 73.71 | 84.74 | 0.02 |
| | | Yes | 96.48 | 99.38 | 40.10 | 57.14 | 0.02 |
| | Adversarial | No | 98.45 | 99.66 | 73.76 | 84.78 | 0.02 |
| | | Yes | **98.38** | **99.66** | **72.58** | **83.99** | **0.02** |

## 6   Conclusions

This work presented a feature selection and consensus process that combined multiple methods, Information Gain, Chi-Squared Test, Recursive Feature Elimination, Mean Absolute Deviation, and Dispersion Ratio, and applied them to multiple network intrusion detection datasets, the original CICIDS2017 dataset, a corrected version of it designated as NewCICIDS, the original HIKARI21 dataset, and an improved version of it designated as NewHIKARI. Several types of ML models, RF, XGB, LGBM, and EBM, were trained with two different feature sets, one with only time-related characteristics and another with more specifically selected relevant features. An adversarial robustness benchmark was performed, analyzing the reliability of the different feature sets and their impact on the susceptibility of the models to adversarial examples.

The employed feature selection process effectively identified the features that contained pertinent information to distinguish between benign and malicious network traffic flows, while discarding the features that did not have enough relevance. For each dataset, the time-related feature set was compared with the more specifically selected features, to guide AI engineers and security researchers to use the most adequate characteristics of network traffic to train their ML models. This reduced number of features can improve computational efficiency, as opposed to using all the available features, and can provide a better robustness, as the models become less overfit to very specific characteristics that would not be usually found in a real computer network.

The adversarial robustness benchmark demonstrated that the first feature set with only time-related characteristics like IAT, idle time, and active time of a flow can provide very good results. Even though the highest overall F1-scores were achieved in combination with the other more specific features, that second feature set sometimes obtained higher false positive rates, which would cause an organization to spend resources and time on unnecessary mitigation measures. Therefore, before deployment, it is pertinent to always evaluate the performance of an ML model for different feature sets, assessing if it exhibits a good generalization to regular traffic and a good robustness to adversarially perturbed traffic, and ensuring a good balance in the trade-off between true positives and false positives.

The best results across several evaluation metrics were achieved in the corrected version of the CICIDS2017 dataset. However, when facing the more recent network traffic flows of the HIKARI21 dataset, the ML models were less robust and exhibited numerous misclassifications. These results suggest that the greater complexity of the more recent cyber-attacks makes it more difficult to distinguish those malicious flows and their corresponding adversarial examples from benign flows that are part of the normal operation of an enterprise computer network. In such cases, adversarial training may not be the best approach, so other defense strategies such as regularization and defensive distillation should be explored.

Overall, by using an improved dataset with more data diversity, selecting the best time-related characteristics and a more specific feature set, and combining it with adversarial training, the ML models were able to achieve a better adversarially robust generalization in the cybersecurity domain. The robustness of the benchmarked models was significantly improved without their generalization to regular traffic flows being affected and without requiring too many computational resources, which enables a reliable detection of suspicious activity and adversarially perturbed traffic flows in enterprise computer networks without costly increases of false alarms.

In the future, it could be valuable to explore the intrinsic explainability capabilities of EBM and experiment with ad hoc explainability methods, to enable a better understanding of the relevance of each feature for each class, and of the reasoning behind each misclassification. To further contribute to adversarial ML research, it is important to benchmark the adversarial robustness of these tree ensembles for multi-class classification and compare them with other types of ML models, including deep learning models. It is also pertinent to explore novel approaches to feature selection that address the relevance of each feature for the robustness of the models, enabling the standardization of the best overall features for cyber-attack classification.

**Author Contributions.** Conceptualization, J.V. and I.P.; methodology, J.V. and M.S.; software, J.V. and M.S.; validation, J.V. and E.M.; investigation, J.V. and M.S.; writing, J.V. and M.S.; supervision, E.M.; project administration, I.P.; funding acquisition, I.P. All authors have read and agreed to the published version of the manuscript.

**Funding.** This work has been supported by the UIDB/00760/2020 and UIDP/00760/2020 projects.

**Data Availability.** Publicly available datasets were analyzed in this work. The original data can be found at: CICIDS2017 (https://www.unb.ca/cic/datasets/ids-2017.html), HIKARI21 (https://doi.org/10.5281/zenodo.4782195). A publicly available method was utilized in this work. The method can be found at: A2PM (https://github.com/vitorinojoao/a2pm).

**Conflicts of Interest.** The authors declare no conflict of interest. The funders had no role in the design of the study; in the collection, analyses, or interpretation of data; in the writing of the manuscript, or in the decision to publish the results.